\documentclass[conference,letterpaper]{IEEEtran}
\usepackage{amssymb}
\usepackage{amsthm}
\usepackage{graphicx}
\usepackage{tabto}

\usepackage[utf8]{inputenc} 
\usepackage[T1]{fontenc}
\usepackage{url}
\usepackage{ifthen}
\usepackage{cite}
\usepackage[cmex10]{amsmath} 
\newtheorem{Def}{Definition}
\newtheorem{Lem}{Lemma}
\usepackage{xcolor}

\begin{document}

\title{Frame Codes For Distributed Coded Computation}
\author{Royee Yosibash}
\date{January 2021}

\author{%
   \IEEEauthorblockN{Royee Yosibash and Ram Zamir}
   \IEEEauthorblockA{EE - Systems Department\\
                     Tel Aviv University, Israel\\
                     Email: royeeyosibash@hotmail.com, 
                     zamir@eng.tau.ac.il
                     }
                     
}

\maketitle
\

\begin{abstract}
    Distributed computation is a framework used to break down a complex computational task into smaller tasks and distributing them among computational nodes. 
    Erasure correction codes have recently been introduced and have become a popular workaround to the well known ``straggling nodes'' problem, in particular, by matching linear coding for linear computation tasks.  
    It was observed that decoding tends to amplify the computation ``noise'', i.e., the numerical errors at the computation nodes.
    We propose taking advantage of the case that more nodes return than minimally required.
    We show how a clever construction of a polynomial code, inspired by recent results on robust frames, can significantly reduce the amplification of noise, and achieves graceful degradation with the number of straggler nodes.
\end{abstract}



\section{Introduction} 
\label{section: Introduction}

In the last years, some algorithms have struggled with the run time of large scale and computationally complex tasks needing many consecutive calculations. A common practice for decreasing run time in such algorithms is using a large distributed system comprised of individual computational nodes.
One of the more significant challenges in these large systems are the "stragglers" – computational nodes whom have a significantly higher response time than their non-straggling counterparts. 
Taking this uncertainty into account calls for a "back-up" scheme in order to ensure high-quality service. One such method is implementing a coding technique taken from the realm of information and code theory. In information theory's terms, a straggler node can be considered as an "erasure" – a symbol in a stream that is 
lost and the receiver knows only through side information that the symbol's real value is unknown. Some of the codes that have been looked into include general maximum distance separable (MDS) codes \cite{LeLaPePaRa2018}, Reed-Solomon (RS) or  Bose–Chaudhuri–Hocquenghem (BCH) codes \cite{RaTaTaDi2019} and the general case polynomial codes \cite{LiAv2020}.
To evaluate which code is best suited for distributed computation, research groups have chosen performance measures such as computational complexity of recovery \cite{LiAv2020}, or average run-time \cite{LeLaPePaRa2018} to show that a certain code is a good solution.
But the above erasure correction codes - when applied to real-valued data - are sensitive to computational errors. In particular, the decoding process may highly amplify these errors due to it being numerically unstable. This amplification of the computational errors incited other codes to be created \cite{FaCa2019, RaTa2019} that try minimizing that amplification effect. However, these codes have decoding schemes that are designed to only use the minimal amount of returning nodes.

In this paper we introduce codes based on results taken from novel ideas from frame theory and random matrix theory. Specifically, we show how new variations of polynomial codes 
could be constructed with these frame-centric design guidelines. 
The noise amplification of these codes follow the theoretical expectations, and the suggested code has near identical amplification as the benchmark for \underline{optimum} noise amplification and also are robust to the number of returning straggler nodes, providing better results as the number of stragglers decreases.

The paper is organized as follows.
Sections~\ref{s.30} and~\ref{section: Coded computing against stragglers} provide background on coded distributed computation 
in the presence of straggling work nodes,
and Section~\ref{section: The noisy setup} proposes a simple noise model due to computation with a finite word-length.
Section~\ref{s.60} describes recent asymptotic results in frame theory, and applies them
to the design of ``irregular'' polynomial computation codes.
Section \ref{s.70} provides numerical results on the noise amplification and condition number of these codes, and shows they outperform the code class defined in \cite{RaTa2019} when the redundancy grows beyond its minimal value.

\section{Distributed Computation} 
\label{s.30}
\begin{figure}[ht]
\center
\includegraphics[width=\columnwidth, height=0.13\textheight]{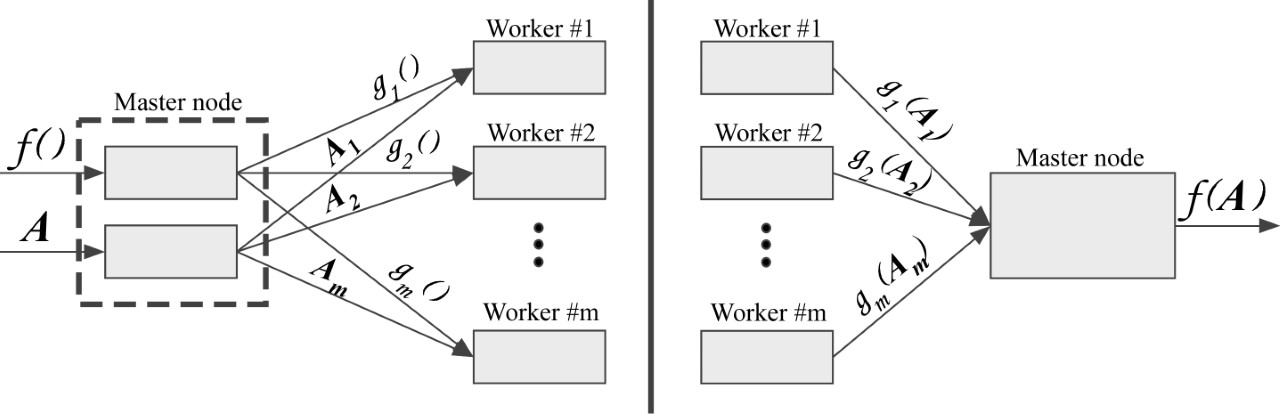}
\caption{The distributed computation setup without coding} \label{fig.10}
\end{figure}

Let some function $f(\mathbf{A})$, with $\mathbf{A}$ some data set which is defined over some arbitrary field $\mathbb{F}^{h \times \ell}$. 
A master node is tasked with computing the function $f(\mathbf{A})$ but it might be too complex and so the task is broken down into some $m$ simpler tasks denoted $g_1,\cdots, g_m$ each operating on $\mathbf{A}$ or a subset of elements in $\mathbf{A}$, denoted also as $\mathbf{A}_1,\cdots, \mathbf{A}_m$. The master node utilizes the worker nodes by sending over the $g_i$ functions with the corresponding subsets $\mathbf{A}_i$ to the nodes so each node $i$ receives $g_i$ and $\mathbf{A}_i$. Each node computes the simpler task $g_i(\mathbf{A}_i)$ and returns the answer to the master node. The restriction on the function set $\{g_i\}$ and the subsets $\{\mathbf{A}_i\}$ is that the master node has to be able to reconstruct $f(\mathbf{A})$ with only $\{g_i(\mathbf{A}_i)\}$. A block diagram of a distributed computation setup is given in Figure \ref{fig.10}.
In this paper we consider a linear function $f$, and specifically matrix-vector multiplication: Here $\mathbf{A}$ is a matrix with $\dim(\mathbf{A})= h \times \ell$ and $\mathbf{x}$ a vector with $\dim(\mathbf{x})= \ell \times 1$. 
The master node uses $m$ computational nodes to calculate $\mathbf{A}^T\mathbf{x}$.
A "na\"ive" approach, that assumes all nodes will return, is to divide the rows of $\mathbf{A}$ into $m$ equal parts creating $m$ matrices that satisfy $\mathbf{A}^T=[\mathbf{A}_1^T,\cdots,\mathbf{A}_m^T]^T$ with each matrix $\mathbf{A}_i$ having $\dim(\mathbf{A}_i)= \lceil \frac{h}{m} \rceil \times \ell$ (adding rows of zeros if $m$ does not divide $h$).
This setup allows the master node to unify the results given by the computational nodes very simply and efficiently: $\mathbf{A}^T\mathbf{x}=[(\mathbf{r}_1)^T,\cdots,(\mathbf{r}_m)^T]^T$.
By assuming the nodes calculate at the same pace and neglecting communication delays it is easy to see the master node has increased the speed of the calculation of $\mathbf{A}^T\mathbf{x}$ by a factor of $m$. In the current setup the "na\"ive" approach seems like the appropriate solution and coding seems unnecessary. The introduction of stragglers is what makes coding an important part of distributed computation.


\section{Coded computing against stragglers}
\label{section: Coded computing against stragglers}

\begin{figure}[ht]
\center
\includegraphics[width=\columnwidth, height=0.13\textheight]{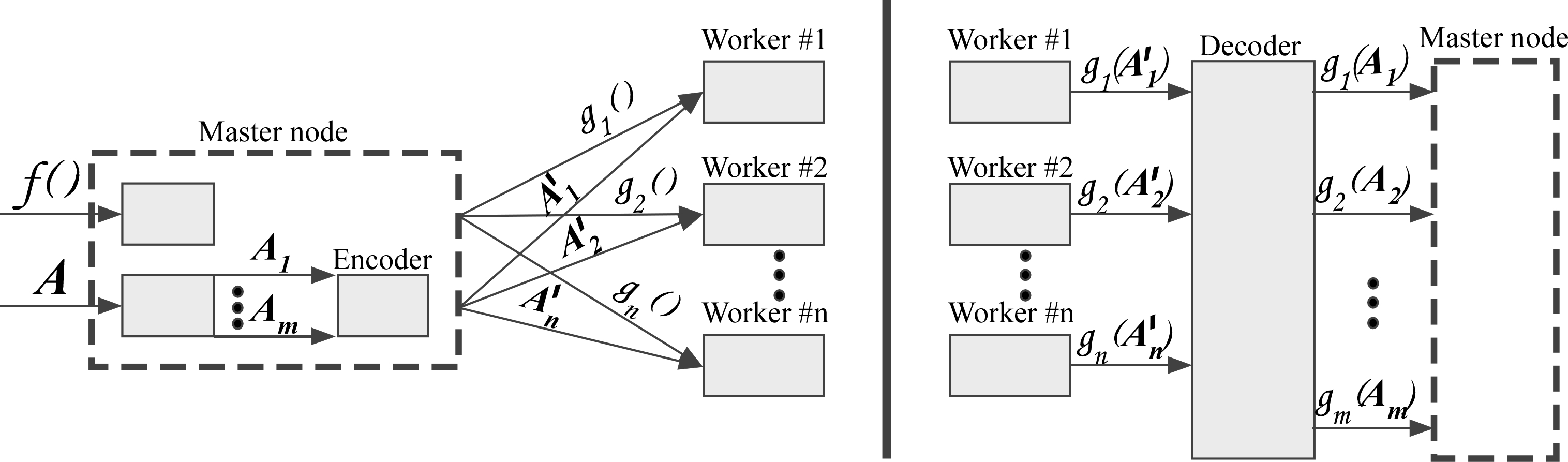}
\caption{The distributed computation setup with coding} \label{fig.11}
\end{figure}

A straggler worker node is the event of a computational node not returning a response in due time (or at all) and the master node deems it unresponsive. This can be seen as the decoder having side information on which of the nodes are valid and which are erased. The number of remaining nodes is denoted as $k$ and we denote the set of node indices that have not been erased as $\mathbf{k} \subseteq [n]$.
In a system with a very large number of nodes the desired runtime, that should have decreased with the number of nodes, is hindered by a lower bound that is the result of the system suffering from having at least one straggler (the probability of which increases with the number of nodes).
As been alluded to before, one way to try to mitigate the straggler problem is by using a coding scheme with erasure resilient codes.
In the same setup described in Section \ref{s.30}, a coding scheme is a function that operates on the $m$ subsets $\{\mathbf{A}_i\}$ to create $n \geq m$ new subsets denoted $\{\mathbf{A}'_i\}$ and $n$ new functions $\{g'_i\}$. These subsets and functions are now sent in the same manner to $n$ nodes. The decoder scheme is a function that operates on the received $\{g'_i(\mathbf{A}'_i)\}$ and converts them back into $\{g_i(\mathbf{A}_i)\}$ for the master node to compute $f(\mathbf{A})$. The encoder might not need all $n$ nodes, and might be able to use only $m\leq k<n$ answers from nodes to retrieve all $\{g_i(\mathbf{A}_i)\}$. In order to simplify, we only discuss cases in which all $g_i$'s are identical and therefore the coding scheme only operates on $\{\mathbf{A}_i\}$. The encoding and decoding schemes slightly alter the setup described in Figure \ref{fig.10} and the new setup is described in Figure \ref{fig.11}.

For a linear function $f$, it is natural to use a linear coding function. Thus the $m$ elements of $\{\mathbf{A}_i\}$ are coded into $n$ elements of $\{\mathbf{A}'_i\}$. This coding scheme can be represented in matrix form by a code generator matrix $\mathbf{F}^T$ with $\dim(\mathbf{F})= n \times m \times \mathbb{F}^{h \times \ell}$:
\begin{equation} \label{e.20}
\mathbf{F}^T[\mathbf{A}_i, \cdots, \mathbf{A}_m]^T = [\mathbf{A}'_i, \cdots, \mathbf{A}'_m]^T
\end{equation}




It is useful to have a representation of erasures as an operator that operates on $\mathbf{F}$. The operator uses the $(n-k)$ elements in $\mathbf{k}'=\{ \alpha \subseteq [n] :  \alpha\notin \mathbf{k}\}$  and nullifies or omits the same index columns in $\mathbf{F}$ (and equivalently the same index rows in $\mathbf{F}^{T}$). This operator is simply a "column retain matrix", $\mathbf{P}$, which right hand multiplies $\mathbf{F}$ to create the equivalent $\mathbf{F}_\mathbf{k}=\mathbf{F\cdot P}$ code generator matrix. 
In the next Sections we will try to compare frames by evaluating, for similar parameters $m,n$ and $k = |\mathbf{k}|$, how the random selection of the $\left(\begin{matrix} n \\ k \end{matrix}\right)$ different subsets $\mathbf{k}$ (uniformly distributed) affect the eigenvalues of these frames.

As mentioned before, the decoder does not necessarily know how many nodes will straggle. Although the decoder knows what is the minimal amount of nodes it requires to calculate the solution it \textbf{might receive more nodes than minimally required}. In the next Section we will discuss how and why a decoder should try to benefit from using more nodes.



\section{The noisy setup} 
\label{section: The noisy setup}

The problem model up to this Section discussed how codes can utilized to mitigate the problem of stragglers. In this Section we will discuss how the existence of computational noise affects the overall numerical instability of coding schemes.
We return to the model of a distributed computation system designed with an encoding matrix chosen as $\mathbf{F}^T$ to better handle stragglers. After transmitting the task to be performed at each computational node, a node $i$ returns the output, $\mathbf{u}_i=\mathbf{A}'_i \mathbf{x}$ with some noise denoted $\mathbf{z}$. While noise from computation might not be detached from the value of $\mathbf{u}_i$, we choose to approximate the noise as some additive i.i.d process in order to simplify the model. The returned transmission is therefore
$\mathbf{r}_i = \mathbf{u}_i + \mathbf{z} = \mathbf{A}'_i \mathbf{x} + \mathbf{z}$

After gathering enough $\mathbf{r}_i$'s the master node can estimate the solution of the computation by, assuming HSNR, using the least-square estimator (using $\mathbf{F}_s$):
\begin{equation} \label{e.40}
\mathbf{E}_{dec} = {\left((\mathbf{F}_\mathbf{k}^T)^\ast\mathbf{F}_\mathbf{k}^T\right)}^{-1} (\mathbf{F}_\mathbf{k}^T)^\ast
\end{equation}

Here $\ast$ is the notation for conjugate transpose.
The {\bf noise amplification} is defined as the MSE divided by the variance of the i.i.d noise. It equivalently defines the degradation in SNR. Therefore, by decoding with the pseudo-inverse of the encoding matrix the noise amplification is:
\begin{equation} \label{e.50}
{\rm Noise \; Amp} = \frac{MSE}{\sigma^2_z}=\frac{1}{k}{\rm trace}((\mathbf{E}_{dec}^\ast\mathbf{E}_{dec})^{-1})
\end{equation}

Because noise amplification is in direct relation to ${\rm trace}(\mathbf{E}_{dec}^\ast\mathbf{E}_{dec})^{-1}$ we see that the noise amplification is also in direct relation to the eigenvalue spectrum of any sub-frame created from the code generator matrix. This leads us to analyze how the code behaves as a {\bf frame}. 


\section{Codes as Frames} 
\label{s.60}

\subsection{Frames and Frame design} 
\label{s.61}
Frame design is, at it's core, the "art" of choosing an over-complete basis $\mathbf{f}_1,..., \mathbf{f}_n$ in some Hilbert space $H$, which satisfies some condition or performs well under some performance measure. In particular, if for some $a,b>0$ the series of $\{\mathbf{f}_i\}_{i\in [n]}$ satisfies
\begin{equation} \label{e.610}
a\| \mathbf{x}\|^2 \leq 
\sum_{i\in [n]} |\langle \mathbf{x},\mathbf{f}_i \rangle|^2  
\leq b\| \mathbf{x}\|^2
\end{equation}

for every $x\in H$ then the sequence is called a {\bf frame}. 
In this paper we deal with vector spaces over the reals or complex. Therefore, when describing a matrix $\mathbf{F}$ that LH multiplies some vector or other matrix (e.g a code generator matrix) we call $\mathbf{F}$ a frame and it's $n$ column vectors (each with $m$ elements) are the sequence $\{\mathbf{f}_i\} _{i\in [n]}$. The "frame aspect ratio" or "redundancy ratio" is defined for a frame $\mathbf{F}$ as $\gamma =\frac{m}{n}$.

In equation \ref{e.610} if $a=b$ then the frame is called a {\bf tight frame} (TF).
A frame is defined as being unit-norm if for all vector norms are $c_{i,i} = 1 ; i\in[n]$. If all $c_{i,i}\neq 0$ then the frame can be transformed into being unit-norm columns by dividing all $\mathbf{f}_{i}$ by $c_{i,i}$.
If a tight frame is also unit-norm then $a=b=\gamma^{-1}$ \cite{Ko2008}.
A frame is an {\bf equiangular frame} (EF) if $|\langle\mathbf{f}_i,\mathbf{f}_j\rangle|$ is a constant for $\{i,j\in [n]; i\neq j\}$. An {\bf equiangular tight frame} (ETF) is a frame which is both a TF and an EF. If a new sub-frame is created by omitting any of the vectors from $\{\mathbf{f}_i\} _{i\in [n]}$ the new frame will have sub-frame aspect ratio of $\beta=\frac{m}{k}$.
As discussed in Section \ref{section: The noisy setup}, we would like to design and analyze frames by their eigenvalue distribution. We choose to have all frames discussed become unit-norm in order to later properly compare them by this performance measure. Going forward, all code generator matrices proposed and discussed, unless specified otherwise, are to have unit-norm columns.
%
As discussed in Section \ref{section: Coded computing against stragglers} and shown in equation \ref{e.50}, our problem is not confined to analyzing a single deterministic frame because the erasure model expands the exploration to frames that have a "goodness" measure that relates to their many possible sub-frames.
This leads us to discuss some known examples from random matrix theory and show good/bad benchmarks to the noise amplification performances.

\subsection{The Random Frame} \label{s.62}
A first natural candidate to discuss is a frame $\mathbf{F}$ with $\dim(\mathbf{F})= m \times n$ whose entries are all random i.i.d Gaussian random variables with variance $\frac{1}{\sqrt{m}}$. Holding the frame ratio constant and $m\rightarrow\infty$ the columns of the matrix have asymptotically a column norm of $1$. By definition, any sub-frame ($k$ of $n$ columns) is also a "random frame". Its been shown in \cite{TuVe2004} that the distribution of the eigenvalues from this frame's Gram matrix converges to the Marchenko–Pastur (MP) density:

\begin{equation} \label{e.1003}
f_{MP}(x) = \frac{\sqrt{(x-\lambda^{MP}_{-})(\lambda^{MP}_{+}-x)}}{2\pi\beta x}\cdot I_{(\lambda^{MP}_{-},\lambda^{MP}_{+})}
\end{equation}
\begin{equation} \label{e.1004}
\lambda^{MP}_{\pm} = \left(1\pm\sqrt{\beta}\right)^2
\end{equation}

As noted in \cite{HaZa2016}, the noise amplification of this type of frame is asymptotically  $\frac{1}{\beta-1}$.

\subsection{ETF's eigenvalue distribution} \label{s.63}
Another natural candidate to discuss are Equiangular Tight Frames due to their rigid column cross correlation requirements. ETFs have suggested to have sub-frame Gram matrix eigenvalues that be asymptotically distributed as in the MANOVA distribution \cite{HaZaGa2018}:
\begin{equation} \label{e.1005}
f_{MANOVA}(x) = \frac{\sqrt{(x-r_{-})(r_{+}-x)}}{2\pi\beta x \cdot (1-\gamma\cdot x)}\cdot I_{(r_{-},r_{+})}
\end{equation}
\begin{equation} \label{e.1006}
r_{\pm} = \left(    \sqrt{1-\gamma\beta} \pm 
                    \sqrt{(1-\gamma)\beta}      \right)^2
\end{equation}

Notice that for $\gamma \rightarrow 0$ the MANOVA distribution converges to the MP distribution.
ETFs seem to have better noise amplification \cite{HaZaGa2018} than most (if not all) frames with the same dimensions and so are a benchmark for good noise amplification. Also, frames that have an eigenvalue distribution that follows the MANOVA distribution also have a condition number that converges to $\kappa = \frac{r_{+}}{r_{-}}$ and shouldn't increase with $n$ as long as $m$ retains the same frame ratio.


\subsection{Cyclic Harmonic Frames} \label{s.64}
Cyclic Harmonic Frames are frames created by choosing some $\mathbf{s}\subseteq [n]$ rows of the $DFT(n)$ matrix \cite{ThHa2017}. These types of frames can have very different noise amplifications after erasures depending on which rows of the DFT matrix are chosen. The first example is choosing $\mathbf{s} = [m]$ (the first $m$ consecutive rows) of the DFT matrix. The pattern created is recognized as the matrix form of a low-pass filter as the truncated $(n-m)$ rows of the $DFT(n)$ matrix are those who give weight to the $(n-m)$ highest frequencies. The band-pass$\backslash$notch filter frame is created in the same way by choosing $\mathbf{s}$ as consecutive series $[n]$ with a cyclic shift over $n$. All these frames have the same Gram matrix because they are all identical up to a scaling factor who is some root of unity (and is canceled out in the Gram matrix) and so have the same noise amplification. These frames also seem to be very noise-amplifying \cite{HaZa2016} \cite{MaZa2013} \cite{SeFe2000}.
We can improve on the LPF by again using the DFT matrix but now choosing the set of rows $\mathbf{s}$ that also qualifies as a difference set. This sub-frame of the DFT was proved to be an ETF \cite{XiZhGi2005}. In \cite{Fa2010} it is shown in that the eigenvalue distribution of a random selection of $\mathbf{s}\subseteq [n]$ converges almost surely to the MANOVA distribution.

\subsection{Polynomial codes represented as frames} \label{s.65}

Now that the motivation for discussing erasure codes and the noise amplification arising from these codes are clear, let us define and discuss the following codes and their generator matrices:

\begin{Def} \label{d.10}
{Polynomial code}
\end{Def}
Given two parameters $(n,m)\in \mathbb{N}$ and two sets $\mathbf{s}\in \mathbb{F}^n$ (here $\mathbb{F}$ is either some Galois field or an infinite field) and $\mathbf{z}\subseteq [n-1]\cup\{0\}$ a polynomial code is a linear transformation of $m$ elements, in the field $\mathbb{F}^{h \times l}$ and denoted $\mathbf{A}_j$, in the following manner:

\begin{equation}
\mathbf{A}'_i =  \sum_{j=1}^{m-1} \mathbf{A}_j \mathbf{s}_i^{z_j}
\end{equation}
Where $\mathbf{A}'_i$ are $n$ encoded elements who are defined over the same field as $\mathbf{A}_i$. The $n$ elements in $\mathbf{s}$ are also called sample points. The $m$ elements of $\mathbf{z}$ are also called the polynomial powers. This linear transformation could also be described in matrix form:
\begin{equation} \label{e.70a}
\left[\begin{matrix}
\mathbf{A}'_1 \\ \vdots \\ \mathbf{A}'_n
\end{matrix}\right]
=
\underbrace{ \left[
\begin{matrix}
s_1^{z_1} & s_1^{z_2} & \cdots & s_1^{z_m}
\\
\vdots &  \vdots &  & \vdots
\\
s_{n-1}^{z_1} & s_{n-1}^{z_2} & \cdots & s_{n-1}^{z_m}
\\
s_n^{z_1} & s_n^{z_2} & \cdots & s_n^{z_m}
\end{matrix}
\right] }_{\mathbf{F}_{PC}^{T}}
\left[\begin{matrix}
\mathbf{A}_1 \\ \vdots \\ \mathbf{A}_m
\end{matrix}\right]
\end{equation}

While polynomial codes are defined over an arbitrary field, we will continue to discuss only polynomial codes defined over the complex field. Its important to note that all samples $\mathbf{s}$ must be distinct and $0\notin\mathbf{s}$. Also, notice that if $\mathbf{z} = [m-1]\cup\{0\}$ then $\mathbf{F}_{PC}$ is the Vandermonde matrix, otherwise it is a generalized Vandermonde matrix \cite{He1929}.
We will now define and discuss a few key examples in this family of polynomial codes: 

\begin{Def} 
\label{d.20}
{Polynomial code with uniform sampling of the unit circle (USPC)}
\end{Def}
We define a polynomial code with a uniform sampling as a polynomial code with $\mathbf{z}$ having $m$ consecutive members in $[n-1]$ and the sample points in $\mathbf{s}$ as:
\begin{equation} \label{e.90}
s_j = \exp\left(\frac{2\pi i}{n}\cdot(j-1)\right) = \omega^{j-1}
\end{equation}

For $\mathbf{z} = [m-1]\cup \{0\}$ these samples create the following code generating matrix:
 \begin{equation} \label{e.100}
     \mathbf{F}_{USPC}^T = \frac{1}{\sqrt{m}} \left[ \begin{matrix}
     1 &  1 &\cdots & 1
     \\
     1 &  \omega^{1} &  \cdots & \omega^{m-1}
     \\
     \vdots & \vdots & & \vdots
     \\
     1 &  \omega^{n-1} &  \cdots & \omega^{(n-1)(m-1)}
     \end{matrix}
     \right]
 \end{equation}

If $\mathbf{z}\neq [m-1]\cup \{0\}$ then in the process of making the frame unit norm it will also divide the columns of the frame by $s_i^{z_1}$ and the unit norm transform of the frame will then be identical to the unit norm transform of $\mathbf{F}_{USPC}^T$.
Notice that frame defined in equation \eqref{e.100} is identical to the $DFT(n)$ matrix with the $(n-m)$ latter columns omitted and then multiplied by $\sqrt{\frac{n}{m}}$ (due to the frame the normalization factor). As mentioned in Section \ref{s.65}, this is recognized as a low-pass and as a frame it has been shown to be very noise amplifying.

\begin{Def} 
\label{d.30}
{Polynomial code with non-uniform sampling of the unit circle (NUSPC)}
\end{Def}
We define polynomial code with a uniform sampling as a polynomial code as a variation of the same code with a uniform sampling. The NUSPC is defined by the parameters $(n,m,b,r)\in \mathbb{N}$ and the set $\mathbf{y}\subseteq [r\cdot b -1]$ with $|\mathbf{y}| = r$. With these parameters the NUSPC sample points set is:
\begin{equation} 
\label{equation: NUSPC sample points}
\mathbf{s}= \{\;\omega^\frac{\mathbf{y}_j + r\cdot b\cdot\alpha}{b\cdot n} :\; \mathbf{y}_j \in\mathbf{y}, \; \alpha\in\mathbb{N}\; \}
\end{equation}

Notice that the constraint on the number of elements in $\mathbf{y}$ is imposed so the number of unique sample points in $\mathbf{s}$ remains $n$. We can also see that a USPC is a special case of a NUSPC and can be created by choosing $y_j$ that are all equally spaced in $r$. This means NUSPC also has the potential to be is very noise amplifying.


\begin{Def} 
\label{d.40}
{Polynomial code with uniform sampling of the unit circle and non consecutive powers.}
\end{Def}
The two codes defined in definitions \ref{d.20} and \ref{d.30} vary the choice in sample points. This choice creates variations/irregularities in the dimension that is subject to erasures. We would also like to try to introduce irregularities in the dimension is not subject to erasures. We define this type of polynomial code as one that uses the same samples as in definition \ref{d.20} but chooses the set $\mathbf{z}$ as one that does not contain a consecutive series in $[n-1]$. 
The discussion in \ref{s.65} implies the following lemma:

\begin{Lem}
\cite{XiZhGi2005} If $\mathbf{z}$ is a difference set, and the samples set are as described in \eqref{e.90}, then this polynomial code is an ETF. 
\end{Lem}

More generally, we expect that even if $\mathbf{z}$ is chosen wisely yet not a difference set (e.g., randomly), the code will still have noise amplification that is close to the good benchmark of ETFs.


\section{Numerical results} \label{s.70}
\subsection{Performance of proposed frame codes} \label{s.71}
\begin{figure} [ht]
\center
\includegraphics[width=\columnwidth, height=0.25\textheight]{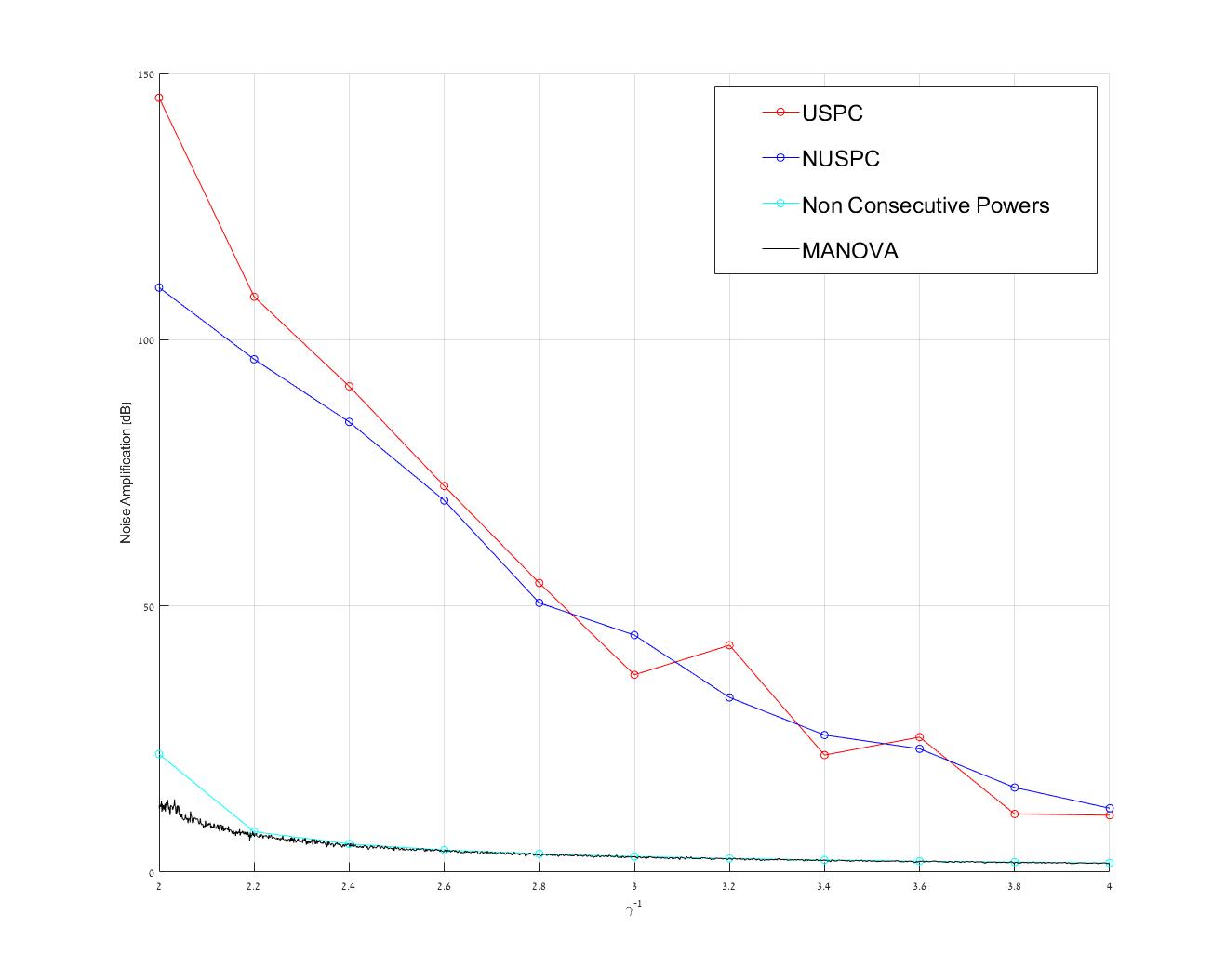}
\caption{Noise amplification vs. $\gamma^{-1}$ for different polynomial codes} \label{fig.30}
\end{figure}

In order to show the performances of the codes described in \ref{s.65}, we show the empirical results for codes with $m=50$ and $n = \gamma\cdot m$ with $\gamma$ in a wide range. The number of nodes retained was set as $\frac{k}{n}=0.5$. Each measurement of a single $m,\gamma$ and code type choose the best (averaged over $10^4$ trials) of 200 valid codes that fitted the description in \ref{d.20}, \ref{d.30}, \ref{d.40}. The best noise amplification of a code with the given set of $m,\gamma$ was then chosen. The comparison is shown in the graph described in Figure \ref{fig.30}.
As theorized in \ref{s.65}, it is clear that the non-consecutive powers polynomial code out-preforms the other codes. Also, the code created by a choosing non consecutive powers closely mimics the noise amplification of a matrix with eigenvalues drawn from the MANOVA probability density function (as expected). In relation to the discussion in \ref{d.40}, we conclude that the introduction of irregularities in the dimension that is not subject to erasures is preferable to only introducing irregularities in the same dimension as the erasures. 

\subsection{Performance of frame codes vs. other numerically stable codes} \label{s.72}

In the works detailed in \cite{FaCa2019, RaTa2019} "orthoMat", "Circ. Perm." and "Rot. Mat." codes are given as examples numerically stable codes (more so than other codes used for distributed computation). These codes choose encoding methods that provide decoding schemes with an upper bound on the condition number of any subset $m\times m $ matrix ($\beta = 1$) because they are meant to used with exactly $m$ returning nodes. As mentioned in Section \ref{section: Coded computing against stragglers}, the master node might receive more nodes than "expected" and can benefit (as shown in Section \ref{section: The noisy setup}) from decoding with more nodes than minimally required. Therefore a robust decoding scheme that can utilize more than $m$ nodes should be preferred. In order to be able to compare our code to other numerically stable codes we must choose codes whom we can modify the original decoding scheme in such a away that retains the original solution for $k=m$ and has better performance for some $k>m$.
We choose to compare our suggested code to the "Circ. Perm." code because the decoding scheme for this code is the inverse of the leftover subset of the generator matrix. Choosing to modify it to be the LS estimator retains the same solution for $\beta=1$ while improving the solution for larger $\beta$.
The comparison between codes is found in table \ref{table1}. $\mathbf{A}$ and $\mathbf{x}$ and other simulation parameters were taken in accordance to those provided in \cite{RaTa2019}. Here $\kappa_{\mathbf{E}_{dec}}$ denotes $\mathbf{E}_{dec}$'s condition number. $\|\cdot\|_F$ is the Frobenius norm of a matrix and $\widehat{\mathbf{A}\mathbf{x}}$ is the estimated $\mathbf{A}\mathbf{x}$.
\begin{table}[ht]
\caption{Code comparison: Non consecutive powers vs. Circ. Perm.} 
\label{table1}
\centering 
\begin{tabular}{c rrrr} 
\hline\hline 
Code Parameters &\multicolumn{4}{c}{$m=29$ \quad $n=31$ \quad $\gamma = 0.935$} \\
Nodes returned &\multicolumn{2}{c}{29} &\multicolumn{2}{c}{30} \\
Code & NCP& CirPerm& NCP& CirPerm \\
\hline 
MSE & $1.85\cdot 10^{-18}$& $1.8\cdot 10^{-18}$ & $2.4\cdot 10^{-19}$& $2.4\cdot 10^{-19}$ \\
$\frac{\|\widehat{\mathbf{A}\mathbf{x}} - \mathbf{A}\mathbf{x}\|_F}{\|\mathbf{A}\mathbf{x}\|_F}$ & 
$1\cdot 10^{-9}$& $1\cdot 10^{-9}$&  $4.9\cdot 10^{-10}$& $4.8\cdot 10^{-10}$ \\
$mean(\kappa_{\mathbf{E}_{dec}})$ & 2.4& 2.3& 3.9& 3.9\\
$min(\kappa_{\mathbf{E}_{dec}})$ & 1.2& 1.2& 3.9& 3.9\\ 
$max(\kappa_{\mathbf{E}_{dec}})$ & 28.9& 55& 3.9& 3.9\\
\hline\hline
\\
\hline \hline  
Code Parameters &\multicolumn{4}{c}{$m=80$ \quad $n=100$ \quad $\gamma = 0.8$} \\
Nodes returned &\multicolumn{2}{c}{90} &\multicolumn{2}{c}{95} \\
Code & NCP& CirPerm& NCP& CirPerm \\
\hline 
MSE & $9.8\cdot 10^{-20}$& $2\cdot 10^{-17}$& $6.3\cdot 10^{-20}$& $2.89\cdot 10^{-19}$\\ 
$\frac{\|\widehat{\mathbf{A}\mathbf{x}} - \mathbf{A}\mathbf{x}\|_F}{\|\mathbf{A}\mathbf{x}\|_F}$ &
$3.1\cdot 10^{-10}$& $9.3\cdot 10^{-10}$& $2.5\cdot 10^{-10}$& $2.9\cdot 10^{-10}$ \\
$mean(\kappa_{\mathbf{E}_{dec}})$ & 5.1& 34.4& 3.3 & 5.1\\
$min(\kappa_{\mathbf{E}_{dec}})$ &  3.6& 2.6 & 2.6 & 2.2\\ 
$max(\kappa_{\mathbf{E}_{dec}})$ & 11.2& $3.6\cdot 10^{4}$& 6.2& 576.6\\
\hline 
\end{tabular}
\label{table:nonlin} 
\end{table}
In table \ref{table1} we see that for $\gamma, \beta$ that are close to $1$ both codes seem to perform almost identically, but if $\gamma$ it "relaxed" and $\beta$ is strictly smaller than $1$ our suggested code outperforms the Circ. Perm. code. For even smaller $\gamma$'s the differences between the codes widen\footnote{For these values of $\gamma$ the testing program warns that some of Circ. Perm's sub-frames are badly conditioned so the comparison wasn't inserted}.

\section{Acknowledgments}
We would like to thank Itzhak Tamo for the insightful discussion on suitable erasure codes for distributed computation systems. This research was partially supported by the Israel Science Foundation, grant \# 2623/20.



\begin{thebibliography}{1}

\bibitem{Fa2010}
B.~Farrell
{\emph Limiting Empirical Singular Value Distribution of Restrictions
of Discrete Fourier Transform.} Journal of Fourier Analysis and Applications 17.4 (2011): 733-753. 

\bibitem{HaZa2016}
M.~Haikin and R.~Zamir 
\emph {Analog coding of a source with erasures}. 2016 IEEE International Symposium on Information Theory (ISIT).

\bibitem{HaZaGa2018}
M.~Haikin, R.~Zamir and M.~Gavish
\emph {Frame moments and welch bound with erasures}. 2018 IEEE International Symposium on Information Theory (ISIT).

\bibitem{He1929}
E. R.~Heineman 
\emph {Generalized vandermonde determinants}. Transactions of the American Mathematical Society 31.3 (1929): 464-476.

\bibitem{LeLaPePaRa2018}
K.~Lee, M.~Lam, R.~Pedarsani, D.~Papailiopoulos and K.~Ramchandran
\emph {Speeding up distributed machine learning using codes}. IEEE Transactions on Information Theory 64.3 (2017): 1514-1529.

\bibitem{LiAv2020}
S.~Li and A. S.~Avestimehr
\emph {Coded Computing: Mitigating Fundamental Bottlenecks in Large-Scale Distributed Computing and Machine Learning}, (2020), pp.\ 66-102.

\bibitem{MaZa2013}
A. Mashiach and R. Zamir
\emph{Noise-shaped quantization for nonuniform sampling} 2013 IEEE International Symposium on Information Theory, Istanbul, 2013, pp. 1187-1191, doi: 10.1109/ISIT.2013.6620414.

\bibitem{RaTaTaDi2019}
N.~Raviv, I.~Tamo, R.~Tandon and A. G.~Dimakis
{\emph Gradient Coding from Cyclic MDS Codes and Expander Graphs}. IEEE Transactions on Information Theory 66.12 (2020): 7475-7489.

\bibitem{SeFe2000}
D. Seidner and M. Feder
\emph{Noise Amplification of Periodic Nonuniform Sampling}, IEEE Trans. Signal Process., vol. 48, no. 1, pp. 275-277, (2000)

\bibitem {TuVe2004}
A. M.~Tulino and S.~Verdú 
\emph{Random matrix theory and wireless communications}. Now Publishers Inc, 2004.

\bibitem{Ko2008}
J. Kovačević and A. Chebira
\emph{An Introduction to Frames}. Foundations and Trends® in Signal Processing: Vol. 2: No. 1, pp 1-94. (2008)

\bibitem{XiZhGi2005}
P. Xia, S. Zhou, and G. B. Giannakis
\emph{Achieving the Welch bound with difference sets}. IEEE Transactions on Information Theory 51.5 (2005): 1900-1907 

\bibitem{ThHa2017}
M. Thill and B. Hassibi
\emph{Low-Coherence Frames From Group Fourier Matrices}. IEEE Transactions on Information Theory, vol. 63, no. 6, pp. 3386-3404, June 2017, doi: 10.1109/TIT.2017.2686420.

\bibitem{FaCa2019}
M. Fahim and V. R. Cadambe
\emph{Numerically Stable Polynomially Coded Computing}. 2019 IEEE International Symposium on Information Theory (ISIT), 2019, pp. 3017-3021.

\bibitem{RaTa2019}
A. Ramamoorthy and L. Tang
\emph{Numerically stable coded matrix computations via circulant and rotation matrix embeddings}. arXiv:1910.06515 (2019)


\end{thebibliography}
\end{document}